\documentstyle[12pt]{article}

\def\pa{\partial}

\def\a{\alpha}
\def\b{\beta}
\def\d{\delta}
\def\ve{{\varepsilon}}

\def\G{\Gamma}
\def\l{\lambda}

\def\s{\sigma}

\def\m{\mu}
\def\n{\nu}

\def\mn{{\mu\nu}}
\def\ab{{\alpha\beta}}
\def\be{\begin{equation}}
\def\bea{\begin{eqnarray}}
\def\ee{\end{equation}}
\def\eea{\end{eqnarray}}

\begin{document}

\hfill BRX TH-575

\vspace{.2in}

\begin{center}
{\large\bf First-order Formalism and Odd-derivative Actions}

S.\ Deser\\Department of Physics, Brandeis University, Waltham, MA
02454
\\ and\\
Lauritsen Laboratory, California Institute of Technology,
Pasadena, CA 91125
\end{center}

{\abstract In this pedagogical  note, we discuss obstacles to the
usual Palatini formulations of gauge and gravity theories in
presence of odd-derivative order, Chern-Simons, terms.}

\section{Introduction}

Writing second order actions in first order form is a particular
case of the general Ostrogradski'i procedure of lowering
derivative order by adding new variables.  This method has proven
particularly illuminating in gauge theories, as exemplified by the
Palatini formulation of general relativity, keeping metric and
connection as independent field variables.  Indeed, it has become
a major activity for generalized gravity models, but it has long
been well-understood for standard GR, where the primary
distinction between the two formulations consists of relative
matter contact terms when spinors are present \cite{weyl}.  In
this note, we consider a more fundamental distinction, quite apart
from matter couplings, in presence of odd-derivative order,
Chern--Simons terms, particularly in D=3 where they play a direct
kinematical role.  We shall see that major changes take place, in
part because gauge invariance can be lost in the Ostrogradski'i
process. We will deal first with the simplest, abelian vector,
topologically massive model (TME), then with the full nonlinear
gravitational one (TMG); both have been exhaustively analyzed in
their second order avatars in \cite{DJT}, whose notations we
follow.

\section{Vectors}

The equivalent Maxwell Lagrangians are
$$
M_2 (A) = -{\textstyle \frac{1}{4}} \: f^2_\mn (A) \;\; , \;\;\;
f_\mn (A) \equiv \pa_\m A_\n -\pa_\n A_\m \; , \eqno{(1{\rm a})}
$$
$$
M_1 (A,F) = -{\textstyle \frac{1}{2}} \: F^\mn f_\mn (A) +
{\textstyle \frac{1}{4}} \: F^2_\mn \; .~~~~~~~~~~~~~~~~~~~~~~~~
\eqno{(1{\rm b})}
$$
Here $F_\mn$ is an independent gauge-invariant variable.  The
Chern--Simons (CS) contributions can also be written in two ways,
$$
C_1 (A) = -{\textstyle \frac{m}{2}} \: \varepsilon^{\mn\a} f_\mn
(A) A_\a \eqno{(2{\rm a})}
$$
$$
C_0 (A,F) = -{\textstyle \frac{m}{2}} \: \varepsilon^{\mn\a} F_\mn
A_\a \; . ~~~~~~~ \eqno{(2{\rm b})}
$$
The ``0$^{\rm th}$ order" form (2b) is no longer gauge invariant
(unlike the integral of (2a)).  [A modification of (1b, 2b) that
we will not consider here, replaces $F_\mn$ by the combination
$f_\mn (B)$, where $B_\m$ is a second potential; this retains
gauge invariance, at the price of keeping derivative order.]  The
combination (1a + 2a) is of course the original TME with a single
massive excitation.  This same model is reproduced by taking (1b +
2a); the CS term does not affect the $F_\mn = f_\mn (A)$ field
equation.  However, (1a + 2b) represents a drastic modification:
we learn that $A_\m$, and consequently $F_\mn$, both vanish; (loss
of gauge invariance of (2b) forces vanishing of $A_\m$).  Finally,
the doubly Palatini variant (1b + 2b) has the field equations
$$
\pa_\m F^\mn + m ^*F^\n = 0 \;\; , \;\;\; ^*\!F^\a \equiv
{\textstyle \frac{1}{2}} \: \varepsilon^{\mn\a} F_\mn \;
.~~~~~~~~~~~~
 \eqno{(3{\rm a})}
$$
$$
F_\mn  = f_\mn (A) + m \:  \varepsilon_\mn   \:^\a\!A_\a \; .
 \eqno{(3{\rm b})}
$$
Upon eliminating $F_\mn$, we find a combined TME/Proca equation,
\be
\pa_\m f^{\m\a} + m \: \varepsilon^{\mn\a} f_\mn - m^2 A^\a = 0 \;
. \ee
 This is known \cite{SDBT} to describe a topological-ordinary mass
 mix, with two massive excitations, of masses $m (\sqrt{2} \pm
 1)$, as against the value $m$ for the single gauge invariant one
 of TME.

 In summary, the differences in kinematic content of the four
 candidate ``TMEs" are extreme, ranging from ordinary TME (in two
 cases) to no excitation to two massive ones, depending on where
 $f_\mn (A)\rightarrow F_\mn$ is inserted.

 \section{Gravity}

\setcounter{equation}{5}

 Let us first list the candidate Lagrangians in the (fully
 nonlinear) gravitational 2+1 TMG of \cite{DJT}.
 $$
 E_2 (g) = \sqrt{-g} \: g^\mn \; R_\mn (g) \eqno{(5{\rm a})}
$$
 $$
 E_1 (g,\G ) = \sqrt{-g} \: g^\mn \; R_\mn (\G )\; .~~ \eqno{(5{\rm b})}
$$
Here $R_\mn (g)$ is the usual metric Ricci tensor, while $R_\mn
(\G )$ is the purely affine version in terms of the (symmetric)
connection $\G^\a_\mn$.  Although $R_\mn (\G )$ is not symmetric
(because $\pa_\m \: \G^\a_{\n\a} \neq \pa_\n \: \G^\a_{\m\a}$),
only its symmetric projection survives in (5b). The equivalence of
(5a) and (5b) rests on the Palatini identity,
 \be
\d \, R_\mn (\G ) \equiv D_\a (\G ) \, \d \, \G^\a_\mn - D_\m (\G
)\, \d \, \G^\a_{\n\a}
 \ee
where $D_\m (\G )$ is the (affine) covariant derivative and $\d
\G$, being the difference of two affinities at a point, transforms
as a tensor.  Varying the affinity in (5b) then implies that $D_\a
(\G ) g^\mn = 0$, determining $\G^\a_\mn$ to be the purely metric
Christoffel symbol $\{ \stackrel{\a}{\scriptstyle \mn} \}$, at
least if $g^\mn$ is invertible \cite{deser}.  Now we adjoin to
(5a,  5b) the two versions of the gravitational CS term. They also
differ from vector CS in that the purely affine version is both
gauge invariant and depends only on the $\G$ and not at all on the
metric:
 $$
 C_1 (\G ) = - {\textstyle \frac{1}{2}} \: \m^{-1}
 \varepsilon^{\l\mn}\:
  \G^\rho_{\l\s}  \Big( \pa_\m \G^\s_{\rho\n} +  {\textstyle
  \frac{2}{3}}\:
 \G^\s_{\m\b}  \G^\b_{\n\rho}  \Big)       \eqno{(7{\rm a})}
$$
 $$
C_3 (g ) \equiv  C_1 (\G^\a_\mn = \{ \stackrel{\a}{\scriptstyle
\mn} \} )\; .
 ~~~~~~~~~~~~~~~~~~~~~~~~
   \eqno{(7{\rm b})}
$$
The $C_1$ version is of first derivative order (in $\G$) while the
metric version (7b) involves three derivatives.  The usual
\cite{DJT} TMG is the sum (5a + 7b), with a single gauge
excitation of mass $\m$.  The combination (5a + 7a), metric
Einstein plus ``Palatini" CS, is--surprisingly--equivalent to pure
Einstein gravity, $R_\mn (g) = 0$, plus a totally decoupled and
rather empty connection sector, with \setcounter{equation}{7}
 \be
 \ve^{\mn\a} \: R^\l_{\s\mn} (\G ) + (\a\l ) = 0
 \ee
where $R^\l_{\s\mn} (\G )$ is the Riemann tensor of $\G$.  This
states essentially that $\G$ is integrable (but still not metric).
The opposite version of Palatini Einstein and metric CS, (5b +
7b), is in fact TMG in disguise: varying the affinity, that only
appears in (5b), tells us that $\G$ is $\G (g)$, whereupon the $\d
g$ equation states that
 \be
 \sqrt{-g} \: G^\mn (g) + \mu^{-1} C^\mn (g) = 0 \;\; , \;\;\;
 C^\mn (g) \equiv \ve^{\m\ab} g_{\b\s} D_\a (R^{\s\n} -{\textstyle
 \frac{1}{4}} \: g^{\s\n} R) \; .
 \ee
So far, then, we have two separate versions of TMG and one of pure
Einstein.  The final, and more difficult, combination is the fully
Palatini model of (5b + 7a).  Since the metric only appears in
(5b), its variation says that $R_{(\mn )} (\G ) = 0$.  Varying $\G
$ gives two terms:
 \be
 D_\a (\G ) (g^\mn \sqrt{-g}\, ) + \Big\{ \ve^{\l\s\n}
 R^\m_{\a\l\s} (\G ) + (\n\m ) \Big\} = 0 \; .
 \ee

This means that the affinity is no longer a metric one, but rather
obeys a far more complicated relation, one we have been unable to
solve. [Of course, a consistent solution is to assume each term in
(10) to vanish separately, which would reduce the system to pure
Einstein.]  Unfortunately, the special fact of D=3 metric
geometry, that metric Riemann and Ricci tensors are equivalent
 \be
 \textstyle{\frac{1}{4}} \: \ve^{\m \ab} \: R_{\ab\l\s} (g)\: \ve^{\l\s\n}
= \det g \: G^\mn (g) \; ,
 \ee
does not at all hold at purely affine level.

In summary, we have extended the TME/TMG vector/tensor D=3 models
to allow for independence of affinity and potential.  In each
case, we have seen quite dramatic differences.  For vectors,
making the higher derivative (Maxwell term) Palatini keeps the TME
structure, while the opposite choice forbids any excitations.
Double Palatini correspond, to  TME + Proca, with two excitations
of different mass.

The gravitational situation is different.  If we let the
``highest-derivative", CS-term only become Palatini, we find pure
metric Einstein gravity.  The other case, in which CS remains
metric just reproduces standard TMG, in analogy to the vector
effect.  Finally, double Palatini here becomes quite involved,
although it does allow for pure Einstein as a consistent solution.

I thank R.\ Jackiw for very useful correspondence, particularly
regarding Palatini TMG. This work was supported in part by NSF
grant PHY04-01667.

\end{document}